\documentclass[usenatbib]{mnras}
\usepackage{graphicx}

\def\neviii{\mbox{Ne\,{\sc viii}}}
\def\neviiia{\mbox{Ne\,{\sc viii}a}}
\def\neviiib{\mbox{Ne\,{\sc viii}b}}
\def\neviiititle{\mbox{Ne\,{VIII}}}
\def\nevii{\mbox{Ne\,{\sc vii}}}
\def\ovi{\mbox{O\,{\sc vi}}}

\usepackage{natbib}
\usepackage{latexsym}
\usepackage{amsmath}

\bibpunct{(}{)}{;}{a}{}{,}

\begin{document} 

\title{Agnostic Stacking of Intergalactic Doublet Absorption: Measuring the  \neviiititle\ Population}

\author[S. Frank et al.]{Stephan Frank$^{1}$\thanks{E-mail: frank@astronomy.ohio-state.edu}, Matthew M. Pieri$^{2}$, Smita Mathur$^{1}$, Charles W. Danforth$^{3}$,\newauthor and J. Michael Shull$^{3}$ \\
$^{1}$The Ohio State University, Department of Astronomy, Columbus, Ohio 43210, USA \\
$^{2}$Aix Marseille Univ, CNRS, LAM, Laboratoire d'Astrophysique de Marseille, Marseille, France \\
$^{3}$CASA, Department of Astrophysical $\&$ Planetary Sciences, University of Colorado, Boulder, CO 80309,USA}

\date{Accepted ? Received ?}

\pagerange{\pageref{firstpage}--\pageref{lastpage}} %\pubyear{2002}

\maketitle

\label{firstpage}

\begin{abstract}

We present a blind search for doublet intergalactic metal absorption with a method dubbed `agnostic stacking'. Using a forward-modelling framework we combine this with direct detections in the literature to measure the overall metal population. We apply this novel approach to the search for \neviii\ absorption in a set of 26 high-quality COS spectra. We probe to an unprecedented low limit  of log N$>$12.3 at 0.47$\leq z \leq$1.34 over a pathlength  $\Delta$z = 7.36. This method selects apparent absorption without requiring knowledge of its source. Stacking this mixed population dilutes doublet features in composite spectra in a deterministic manner, allowing us to measure the proportion corresponding to \neviii\ absorption. 
We stack potential \neviii\ absorption in two regimes:   absorption too weak to be significant in direct line studies  (12.3 $<$ log N $<$ 13.7), and strong absorbers (log N $>$ 13.7). We do not detect \neviii\ absorption in either regime. Combining our measurements with direct detections, we find that the \neviii\ population is reproduced with a power law column density distribution function with slope $\beta = -1.86  \substack{+0.18 \\ -0.26}$ and normalisation log $f_{13.7} = -13.99 \substack{+0.20 \\ -0.23}$, leading  to an incidence rate of strong \neviii\ absorbers  $dn/dz =1.38 \substack{+0.97 \\ -0.82}$.
We infer a cosmic mass density for \neviii\ gas with 12.3 $<$ log N $<$ 15.0  of  $\Omega  _{\textrm{\neviii}} = 2.2 \substack{+1.6 \\ _-1.2} \times 10^{-8}$, a value significantly lower that than predicted by recent simulations. We translate this density into an estimate of the baryon density $\Omega _{b} \approx 1.8 \times 10^{-3}$, constituting 4\% of the total baryonic mass. 
\end{abstract}
 

\section{Introduction}\label{introduction}

Galaxies do not exist in a vacuum. They are embedded in a complex, large-scale structure of gas and dark matter, often called the `cosmic web'. Not only during their formation, but throughout their whole lifetime they exchange mass, energy, momentum, angular momentum, and entropy with their surroundings in the form of in- and outflows as well as radiative processes and gravitational interaction. Thus, determining the physical state of the gas  in the intergalactic medium (IGM) and the subset of gas in the vicinity of galaxies (the circum-galactic medium; CGM)  can yield key insights into the mechanisms that govern the interaction between the cosmic web and galaxies. These properties include the density, temperature, metallicity, clumping, ionization, and environment with respect to local galaxies and the broader cosmic web. 

Most observational information on the IGM and CGM is derived from the study of quasar absorption lines. The Lyman-$\alpha$ transition traces residual neutral hydrogen in post-reionization universe, is immensely useful because it is common and because it is a good indicator of overdensity \citep{Rauch1997}.  Metal absorption also plays an important role. The ionisation status of a metal absorber is directly influenced by its density, temperature, metallicity, and the radiation field in which it bathes (e.g. \citealt{Pierietal2014}). When temperatures exceed T$=10^{5}${} K the gas becomes highly ionized and Lyman-$\alpha$ lines (along with several others) become disfavoured. This makes the study of such `warm-hot' gas challenging. At $z>2$ the majority of barons are readily observable in the Lyman-$\alpha$ forest, but at lower redshifts significant quantities of baryons become heated chiefly due to gravitational shocks  \citep{Cen1999, Dave2001}. The emerging dominance of the Warm-Hot Intergalactic Medium (WHIM) combined with the observing challenge noted above is thought to lead to the so-called 'missing baryon problem' at the present epoch \citep{Fukugita2004}. 

Alternatives to quasar absorption have been explored to by-pass these observational challenges. Cross correlations of CMB anisotropy and both galaxy distributions \citep{Genova2015} and weak lensing maps \citep{Atrio2017} show broad consistency with expected WHIM gas populations, while attempts have been unsuccessful for detection in the diffuse X-ray background \citep{Cappelluti2012, Roncarelli2012}. Recent results indicate that a significant proportion of baryons have been detected in filaments between SDSS DR12 galaxies \citep{Alam2015}{} through stacking of galaxy pairs and measurement of an associated thermal Sunyaev-Zeldovich excess\citep{deGraff2017, Tanimura2017}.

While detecting the warm-hot IGM in X-rays has been challenging \citep{Nicastro2002,  Mathur2003, Nicastro2016}, the warm-hot CGM is well studied in X-rays (see e.g. \citealt{Gupta2014, Gupta2017, Nicastro2017}). Although the warm-hot gas is hence detectable in the X-rays (see \cite{Bregman2007} for a complete overview), the effort to characterise warm circum- and intergalactic gas at redshifts z$<$1.5 has largely focused on \ovi, five-time ionized oxygen, in the UV.   The \ovi\ doublet (1031.9 \AA\ and 1037.6 \AA) samples gas with temperature $2 \times 10^{5} K <T< 6\times  10^{5} K$ in collisional ionization equilibrium \citep{Gnat2007}, but traces gas over a wide range of environments, spanning from the local ISM to the IGM (see e.g. \cite{Jenkins1978, Tripp2000, Howk2002, Danforth2005, Tumlinson2011} and references therein).  \ovi\ can also arise, however, in photoionized gas extending the temperature range traced down to T$\sim 10^{4}$K, which is typically the case at $z>2$ (e.g. \citealt{Aguirreetal2008}).
Hence, the interpretation of \ovi\ detections is debated \citep{Danforth2008, Thom2008, Tripp2008, Oppenheimer2009,  Smith2011, TepperGarcia2011, Cen2012}.
Broad Lyman-$\alpha$ absorbers (BLAs, 
for a definition see \cite{Richter2006}) have been considered 
as a solution to the detection of this challenging gas phase
\citep{Richter2004, Sembach2004,  Williger2006, Lehner2007, Danforth2010, TepperGarcia2012}. However, BLAs are difficult to identify in UV spectra because they are expected to be shallow ($\tau_{0}$ (H I)$<0.1$) and are challenging to distinguish from blends of narrower lines.

Given these limitations, alternatives to \ovi\ and BLAs must be considered, and one promising tracer is seven-times ionized neon (\neviii). Neon is the 5th most abundant element in the universe, and as such represents a potential tracer of intergalactic (IGM) and circumgalactic (CGM) gas. In particular \neviii\ unambiguously traces gas warmed to  $4\times10^{5} K < T < 2 \times 10^{6} K $ through collisional ionization (\cite{Gnat2007}, inside this temperature range the collisional ionization equilibrium (CIE) fraction of \neviii / Ne is above 0.01), due to the high  ionization potential of \nevii\ (207 eV). Photoionization models, however, for even low-density \neviii\ gas are highly uncertain due to the lack of good constraints on the soft X-ray- and EUV parts of the metagalactic background, see e.g. \cite{Shull2014}.
\neviii\ provides a strong absorption doublet in the UV at 770.409\AA\ (in the following \neviiia) and 780.327\AA (\neviiib), which at the low redshifts of interest must be measured in the observed frame UV, and hence must be observed from space.

The Cosmic Origins Spectrograph (COS, \citealt{Green1998, Green2012}), is an ultraviolet (UV) instrument aboard  the Hubble Space Telescope (HST). It has increased sensitivity compared to previous instruments
such as the Far Ultraviolet Spectroscopic Explorer (FUSE) or the Space Telescope Imaging Spectrograph (STIS), and has thus enabled high-sensitivity, medium resolution spectroscopy of faint astronomical objects in the wavelength range between 1135 \AA\ and 3200 \AA{}\footnote{In the special G130M 'super-blue' setting it is possible to reach as low as 1050 \AA.}. It is therefore ideally suited for the search of \neviii\ absorption signatures at low redshifts. Given the doublet's restframe wavelengths, redshifts $z>$0.5 are needed to shift the \neviii\ absorbers into the COS bandpasses. 

Unambiguous identification of the doublet signature of \neviii\ is challenging since it may be hampered by blends with unrelated absorbers (especially towards higher redshifts where the forest becomes thicker) and it is reliant on the availability of additional confirming absorption species. Exploiting even the best available UV-spectra has lead thus far only to 10 direct detections of \neviii\ absorbers \citep{Savage2005, Savage2011, Narayanan2009, Narayanan2011, Narayanan2012, Tripp2011, Meiring2013, Hussain2015, Qu2016, Pachat2017}. All of the unambiguously detected individual systems have column densities log N $>$ 13.0.\footnote{Here and in the following, the unit for column density N is [cm$^{-2}$].} Simulations predict the bulk of the \neviii\ absorber population to lie between 12.0 $<$ log N $<$ 12.7 \citep{TepperGarcia2013, Oppenheimer2013, Rahmati2015}. Hence, we chose to relax the requirement of unambiguous detection and search for the weak distributed signal of \neviii\ at the noise limit of these spectra. Such methods were initially developed as pixel correlation based searches \citep{Cowieetal1998, Daveetal1998, Aguirre2002, Pierietal2010a} and explored further through the use of absorber-frame spectral stacking \citep{Pierietal2010b, Pierietal2014}. We also supplement this search for weak absorption, by applying our analysis methods to the search for strong \neviii\ detectable on an individual basis. In this way we take advantage of our blind absorption population analysis techniques to attempt an alternative measurement of the \neviii\ column density distribution function.

Here we refine a technique initially presented in \citet{Pieri2014}, developed to probe for doublet absorption with a blind statistical approach. This method relaxes the requirement that absorbers be statistically significant detections individually, opening a more numerous population hidden to traditional line searches. It also circumvents potential subjectivity of individual line identifications and automatically takes into account chance pairings of lines which reproduce the exact \neviii\ doublet ratio. \citet{Pieri2014} only sought to explore the potential signal strength of triply ionized carbon, here we attempt to measure a physically motivated metal absorber population via a power-law parametrisation of the column density distribution function (CDDF) and suites of phenomenological mock spectra. We limit our analysis to a range of power-law slopes, -3.6 $\leq \beta \leq$ -1.3.  

This paper is structured as follows: after giving an overview of the dataset used in our analysis, and initial data preparation beyond the standard data reduction (Section \ref{Data}), we proceed to introduce the idea of an `agnostic' stacking procedure (Section \ref{idea_agnostic}), detailing both the necessary modelling of the underlying absorber populations and the subsequent optimal choice for selecting apparent absorption. After summarising our results in Section \ref{Results}, we conclude with the discussion in Section \ref{Discussion}.

Throughout the paper we adopt the cosmological parameters of the  \citet{Planck2015}, i.e. h=0.678$\pm$0.009, $\Omega_{m}$=0.308$\pm$0.012, $\Omega_{b}$=0.0485$\pm$0.0005. 

% *******************************************************************
% *******************************************************************

\section{Data}\label{Data}

We exploit a homogeneous set of UV-bright QSO absorption spectra observed with COS. For details of the sample and data reduction see \citet{Danforth2016}. We infer the transmitted flux $F=f/C$, where $f$ is the observed QSO flux and $C$ is the unabsorbed QSO continuum, via  a semi-automated continuum-fitting technique
developed for optical SDSS spectra \citep{Pierietal2010b, Pieri2014} and adapted for use in higher-resolution FUV data. While the full details of this process can be reviewed in \citet{Danforth2016},  we briefly mention the salient points here. First, spectra are split into small segments (5-15 \AA\ width). Continuum pixels within each segment are identified as those for which the flux divided by the error is less than a 1.5$\sigma$ threshold below the median flux/error value. Thus, absorbed pixels (flux
lower than the segment median) are excluded, together with regions of increased noise (error higher than segment average). This process is iterated until minimal change occurs in the population of continuum pixels, or until only 10 percent of the original pixels in the segment remain classified as continuum. The median value of those continuum pixels is then recorded as a continuum flux node for the segment, and a spline function fitted between the nodes. The continuum fit of each spectrum was then checked manually and adjusted as needed.

Hereafter we refer to the transmitted flux, F, simply as `flux' and the error estimate $\sigma$ as the error estimate of the transmitted flux simply as `error esimate'. In total, 26 sightlines are suitable for our search program.
In this section we list basic characteristics of the data pertaining directly to the our search, namely the noise characteristics and pathlength. Additionally, we summarise the effects of rebinning the spectra in order to maximise detection efficiency.

\begin{figure}
 \includegraphics[width=90mm]{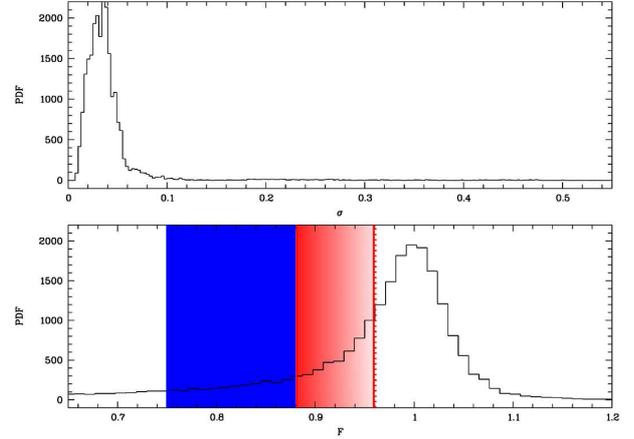}
 \caption{Distribution of transmitted flux ({\it lower panel}) and its error estimate ({\it upper panel}) in each rebinned pixel after rebinning. Note that the majority of the pixels now have a $\sigma < 0.05$ and so S/N per pixel $>$20. The blue area delineates the strong absorber search, the red the search for weak absorbers.  Note that there is no hard criterion for the upper flux limit in the weak absorber search, the dotted line indicates the approximate position of a pixel with a median S/N of 25. Note also that the distribution of flux is not symmetric around $F=1$, with more pixels on the lower side, indicating the presence of absorption at all levels. Exploiting properties of the source of this asymmetry in the context of strong noise is the basis of our search for weak \neviii.}
 \label{flux_and_noise_pdf}
\end{figure}

We select QSOs with an emission redshift $z_{em} >${} 0.5 in order to provide coverage of  \neviiia\ in the COS G160M grating.
Table \ref{tab:sightlines}{} gives an overview of the available COS-archival spectra with the available redshift path, and an estimate for the average S/N in the range suitable for \neviii\ detection. For some of the sightlines, all of the absorption features (above a certain significance threshold) have been securely identified, whereas for the majority of the high redshift ($z_{em} > 0.8$) QSOs those absorbers remain mostly unidentified, yet significant \citep{Danforth2016}.

In all spectra, we mask out problematic wavelength ranges, such as Galactic Ly$\alpha${} absorption or a small window around 1307 \AA, severely affected by the removal of emission lines. Furthermore, we set the maximum wavelength to be at least $\Delta v > 5000$ km/s blueward of \neviii\ in the QSO restframe, to avoid intrinsic or associated absorption. 
This leads to pathlengths for the search for \neviiia\ as listed in Table \ref{tab:sightlines} for each sightline.
The total pathlength suitable for \neviii\ detection summed over the 26 sightlines is $\Delta z = 7.36$, the median (mean) redshift for \neviii\ is 0.63(0.88), and the absorber redshift spans the a range from 0.47 $\leq$ z$_{abs} \leq 1.34$.
 
%******************************************************************************* 

\subsection{Supplementary data preparation}

The COS G160M grating affords us with a spectral resolution of $\lambda / \Delta \lambda \sim$18,000 (equating to $\Delta v \sim 17$ km/s), and an initial wavelength sampling of  $\sim 10^{-2}$\AA, translating into a velocity separation of $\Delta v \sim 2$ km/s for each pixel. We expect the signatures of \neviii\ absorption to be significantly broader than this separation,  since we want to focus on absorbers residing at temperatures T$>10^{5}$K. Hence, in order to maximise the signal-to-noise ratio (S/N) per pixel (without significantly reducing the desired signal), we aggressively rebin our spectra without losing information. Note that this is consistent with our goal to {\it detect}{} \neviii\ absorption, and not to infer velocity structure of such absorbers.

For a homogeneous selection function it is desirable to fix the velocity of our rebinned sample and this is achieved with a logarithmic binning. This also offers the desirable feature of setting a fixed wavelength solution in the absorber frame composite spectrum irrespective of absorber redshift. We maximise the \neviiib\ doublet signal by requiring an integer number of wavelength bins between the doublet lines. The combination of these drivers results in a wavelength solution of 
\begin{equation}
\Delta {\log} \lambda = \frac {1}{56} \log \Big[ \frac {\lambda_{\rm \neviiib}} {\lambda_{\rm \neviiia}}\Big] = 9.9202 \times 10^{-5}, 
\end{equation}
where the \neviii\ doublet restframe wavelengths are 770.409 \AA\ (\neviiia) and 780.327 \AA\ (\neviiib) as quoted in \citet{morton2006}{}. Translated into a velocity separation this becomes $\Delta v=69$ km/s. The full width at half-maximum (FWHM) for \neviii\ absorbers of temperature T are related by $(FWHM / 15.1\  {\rm km/s})  =  \sqrt{T/10^5 K}$. Therefore our binning of $\Delta$v=69 km/s  is well-matched to the expected temperature of \neviii\ and sufficiently wide to ensure that almost all of a putative absorption feature will fall in just one resel. A doubling or halving of the line width leads to only a 20\% change in the detection significance and lines which group in complexes will generate signal irrespective of whether they fall in the same bin or neighbouring bins.
Figure \ref{flux_and_noise_pdf} shows the distribution of the error in the flux after rebinning (upper panel), and the resulting pixel flux distribution in the window relevant to our search. The majority of the rebinned pixels have a error estimate of the flux between 0.02 and 0.04. The probability distribution function (PDF) of the flux shows a strong asymmetry around the flux of unity, where the drop-off towards flux above the continuum is much faster. This indicates the presence of absorption prevalent even at a low flux decrement. It is the very basis of our stacking approach for the weak absorbers to exploit this asymmetry in the regime that is dominated by noise.

\begin{figure}
\includegraphics[width=90mm]{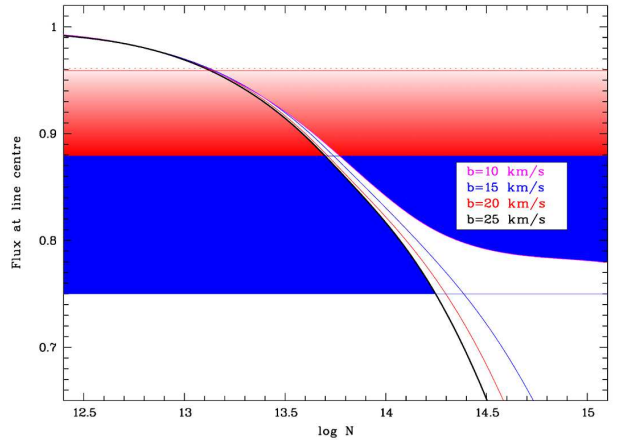}\caption{The minimum flux (i.e. at line center) for the \neviiia\ component  with a given column density, assuming different b  parameters, and given the pixel size after rebinning. Note that when probing the column density regime for \neviiia\ below the limit of traditional direct line searches (log N $<$ 13.7), we expect the feature to create flux decrements lower than $\sim$10\%, hence the need both for data of high-S/N and a statistical analysis rather than direct detection. The coloured regions indicate the flux boundaries for the two different search regimes as in Fig. 1.} 
\label{logN_vs_minflux}
\end{figure}

Fig. \ref{logN_vs_minflux}{} shows the expected minimal flux at line centre for different choices of the absorbing column density log N$_{\rm NeVIII}$ given our rebinning. Here, we assume Voigt profiles with a single velocity component and varying broadening parameters. It is clear that trying to reach column densities lower than the current observational limit from direct detection efforts (log N$_{\rm NeVIII} \sim$13.7), we need to sample pixels affected by less than 10$\%$ absorption, indicating the importance of high S/N.

We do not attempt to remove pixels that have been identified to be part of known non-\neviii\ absorbers, since such identification is incomplete for our sample (especially for sightlines with high QSO emission redshifts). Additionally, our analysis includes the presence of such absorbers and is dealt with by our heuristic mocks (see Section~\ref{mocks}). 

\begin{table*}
\caption{COS-archival QSO sightlines suitable for the search for \neviii\ absorption. The naming convention is the same is in Danforth 2016, where all relevant information on the initial data reduction can be found. The total available pathlength for \neviiia\ takes into account masking of areas of poor S/N or problematic wavelength regimes (such as Galactic Ly$\alpha$ absorption), and an additional filter excluding all pixels outside the flux regime expected for \neviii\ absorption: F $<0.88${} for the weak search, F $<0.75${} for the strong search. The number of available pixels and pathlengths for the strong absorber search with its lower flux cutoff is on average $\sim 10\%${} higher. The S/N quoted for each sightline is the median for all pixels in the regime suitable for the weak \neviii\ search. Note that the number of pixels is given after our additional rebinning. The references in the comments column indicate the direct detection of \nevii\ absorbers along those sightlines.}
\label{tab:sightlines}
\begin{tabular}{@{}lccccc}
  Name        & Emission redshift z$_{em}$ & $\Delta z _{\neviiia}$ & Number of pixels &  Median S/N per pixel & Comments \\
\hline
pks0405   & 0.574 & 0.0707 & 203  & 91.7 & \citet{Narayanan2011} \\
pg1424    & 0.610 & 0.0583 & 167  & 23.9 & \\
he0238    & 0.631 & 0.0766 & 219  & 36.9 & \\
3c263     & 0.646 & 0.0913 & 258  & 43.9 & \citet{Narayanan2009, Narayanan2012} \\
pks0637   & 0.650 & 0.0826 & 232  & 29.6 & \\
s080908   & 0.656338 & 0.0949 & 267 & 21.1 & \citet{Pachat2017} \\
3c57      & 0.670527 & 0.1059 & 296 & 35.0 & \\
pks0552   & 0.68000 &  0.1031 & 288 & 29.2 & \\
sbs0957   & 0.746236 & 0.0893 & 233 & 16.1 & \\
sbs1108   & 0.766619 & 0.1133 & 306 & 5.2 &  \\
s234500   & 0.7789429 & 0.2042 & 544 & 26.9 & A.k.a. pmn2345 \\  
sbs1122   & 0.852 & 0.2336 & 544 & 19.9 &  \citet{Pachat2017} \\
f0751     & 0.915 & 0.2793 & 709 & 35.0 & \\ 
pg1407    & 0.940 & 0.3164 & 799 & 66.1 & \citet{Hussain2015} \\
q0107     & 0.956 & 0.3295 & 834 & 26.1 & \\
l01070b   & 0.957039 & 0.3169 & 804 & 25.0 & \\
pg1148    & 0.975  & 0.3535 & 882 & 50.3 & 3 absorbers, \citet{Meiring2013} \\
he0439    & 1.053  & 0.4087 & 1012 & 29.1 & \\
s100535   & 1.0809  & 0.4356 & 1064 & 23.9 & \\ 
f020930   & 1.128   & 0.3811 & 900 & 32.8 & \\
pg1206    & 1.16254 & 0.3668 & 873 & 32.5 & \citet{Tripp2011} \\
pg1338    & 1.21422 & 0.3732 & 876 & 33.1 & \\
l14350    & 1.30791 & 0.6435 & 1479 & 62.6 & \citet{Qu2016} \\ 
pg1522    & 1.32785 & 0.5855 & 1340 & 52.1 & \\
q0232     & 1.437368 & 0.6531 & 1485 & 36.3 & \\
pg1630    & 1.47607  & 0.5904 & 1329 & 39.3 & \\
\hline
          & Total & Path $\Delta$z=7.3554 & N$_{pixels}$=18011 & & \\
\hline
\end{tabular}
\end{table*}

% *********************************************************************
% *********************************************************************

\section{Agnostic stacking and constraints using mock data}\label{idea_agnostic}
\label{bigpicture}

In the general framework of spectral stacking, typically spectra of well-defined homogeneous samples are co-added in the hope of improving the S/N of the resulting composite spectrum over the often low-quality individual spectra, such that salient features of the overall object population can be discerned. As demonstrated in \cite{Pierietal2010b} and \cite{Pierietal2014}, one can remove the requirement that the entire spectrum characterises the object population, since other uncorrelated absorption in the spectrum is simply another source of noise in the composite spectrum (once broad depressions in the spectrum are corrected for). Here, we propose to go a step further by relieving ourselves of the need to identify a clean object absorber sample, constructing what we dub an `agnostic stack'. Agnostic stacking must be forward modelled in order to compare its results with potential absorber populations. 
Furthermore, we utilise these mocks  to optimise arbitrary analysis choices in order to maximise the expected signal (see Section~\ref{weaksearch}).

We select all pixels showing apparent absorption regardless of their source transition (hence the term `agnostic'), and treat {\bf all} of these selected pixels as putative \neviiia\ absorbers. 
We then shift each of these \neviiia\ absorber candidates into the \neviiia\ rest-frame along with the rest of the spectrum with the potential doublet signal but also additional unrelated absorption . Although we are obviously primarily interested in probing wavelength separations required to cover the rest-frame  \neviiib\ feature, spanning a wider wavelength range can serve as important resource to diagnose eventual artifacts like the presence of uncorrelated absorption, that may mimick our doublet signal. For example, the wavelength ratio of Ly$\delta${} to Ly$\epsilon${} is very close to the \neviiia{} to \neviiib{} ratio. In order to assess whether a potential signal could be due to the presence of such Lyman series lines, we would look hence at the location of the lower order transitions Ly$\alpha$, Ly$\beta$, and Ly$\gamma$, which are expected to produce a stronger feature. For the details of the subsequent spectral stacking, we refer the reader to Section 3.4.  If a significant proportion of the selected sample is indeed \neviiia\ absorption of sufficient strength, then we expect the resulting composite spectrum to show a signal at the restframe wavelength of \neviiib. This signal is diluted by both observing noise and non-\neviii\ absorbers. Adding  non-\neviii\ absorbers increases both the raw composite spectral noise and the signal dilution factor. The crucial question then becomes what is the optimal absorption selection that maximises the potential \neviii\ S/N in the competition between composite noise reduction and increasing signal dilution.

In order to address this question (and optimise the analysis) it is necessary to forward model the selection and stacking procedure by creating and testing mock spectra. The following section describes details of the generation of such mock spectra, and the procedure to estimate the expected signal strength and its significance with their help.

What we arrive at is an end-to-end analysis comparison of the \neviiib\ signal in the observed composite spectrum to that found in mock composite spectra. This is combined with a measurement of the flux uncertainty in the composite spectrum. Thus we constrain viable \neviii\ populations.

\subsection{Generation of mock spectra}
\label{mocks}

We compute mock spectra that match the  pathlength and wavelength logarithmic binning as the observed data. In a first step, we create random realisations of \neviii\ absorbers from model column density distribution functions to populate those spectra. We assume the column density distribution function (CDDF) to be represented by a single power-law over the column density regime we want to explore
\begin{equation}
 f(\neviii) = \partial ^{2} \mathcal{N} / \partial X \partial N _{\textrm{\neviii}} \sim \textrm{f}_{0} (\frac{N}{N_{0}})^{\beta},
 \end{equation}
  where $\mathcal{N}$ is the number of absorbers per unit column density N$_{\textrm{\neviii}}$ and unit absorption distance\footnote{The absorption distance is given by \\dX$ = (1+z)^{2}[\Omega_{m}(1+z)^{3}+\Omega_{\Lambda}]^{-\frac{1}{2}}$dz.} X, f$_{0}$ is a normalisation at a given column density N$_{0}$, and $\beta$ the slope of the CDDF. In the following we will be using log N$_{0}$ = 13.7 as our reference column density, because it marks the approximate delineator below which individual line searches become impossible due to the S/N constraints of the data, and use the notation $f_0 = f_{13.7}$. Integrating over all column densities probed by us yields the line density $d\mathcal{N}/dz = \int \int f\textrm{(\neviii)} dNdX$. We place these absorbers in the spectra by randomly assigning them a redshift.\footnote{Note that hence we do not take into account clustering, or redshift evolution, both of which we deem secondary effects not significantly altering our selection process.} In addition to the randomisation in column density and redshift, we also assign each absorber a broadening parameter $b$, randomly drawn from a flat distribution between 15 and 40 km/s.

\begin{figure}
\includegraphics[width=90mm]{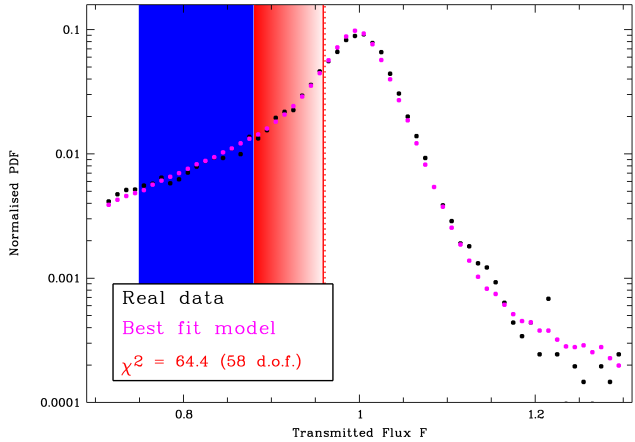}\caption{An example of fitting the non-\neviii\ absorption. Shown as black points is the PDF of the flux for real data. The magenta points show the PDF for a particular set of mock spectra after all three components have been added (\neviii\ absorption ; additional non-\neviii\ absorption ; and the appropriate noise). We adjust the model for the non-\neviii\ absorbers to minimise the reduced $\chi ^2$. The blue and red areas delineate the search windows for strong and weak absorbers as in Figs. 1 and 2.}
\label{normalised_pdf_mock_vs_real}
\end{figure}

This population of \neviii\ absorbers is mixed in with a larger distribution of non-\neviii\ absorbers, both weak and strong, as well as the noise distribution provided by the data. Hence, we must add additional absorption and noise. Here we are following a purely heuristic modelling scheme. Each pixel's flux is diminished by multiplying it with a factor$P(\tau) = A \exp (-c \tau)$, where A and c are fit to the data, as detailed below.\footnote{This proved to be  more effective than the more standard log-normal opacity distribution.}
Additionally, we add noise drawn from the noise PDF of the real data. In essence, we transfer the error estimate array from the real data pixelwise onto the array for the mock spectra, and then for each pixel we 
generate a random realisation of noise from a 
Gaussian distribution with a width given by that error estimate. Note that in this way, the specific noise characteristics of each spectrum are retained in the structure of the mocks, i.e. adjacent pixels in the mock remain affected by similar noise. Also note that while we are aware of effects of non-Gaussianity for the COS noise estimators, we deem those to be non-problematic for our mocks, since they will not generate artifacts on the scale of the \neviii{} doublet spacing, and are being modelled by having the heuristic model PDF being a good fit to the data.

As illustrated in Fig.~\ref{normalised_pdf_mock_vs_real}, the parameters of the absorption model are adjusted such that a best fit for the pixel flux distribution with all three components (\neviii\ absorbers, additional non-\neviii\ absorption, and noise) is achieved when compared to the real dataset. Note that we include only pixels whose flux is greater than F$\geq$0.65 when calculating the goodness of fit, as we do not try to reproduce the distribution towards lower fluxes, where pixels are affected by saturation and deemed to be too absorbed for harbouring the relatively weak \neviii\ absorption for which we search here. Hence, we do not expect adding the additional absorption to adequately fit the flux pdf below that value - and indeed, we generally end up with too few pixels showing near zero fluxes, indicating the strongest possible absorption.  

We generate a sufficient number of suites of mock spectra such that the statistical sample variance is negligible with respect to the noise estimate from the real data. Typically this results in creating on the order of 100 model spectral data-sets for each combination of the two relevant free parameters (slope $\beta$ and normalisation f$_{13.7}$ of the CDDF). 

This modelling automatically accounts for the existence of unrelated absorbers, and furthermore describes any significant sample of pixels effected by fixed pattern noise from the COS CCD (see \citet{Keeney2012} for an overview of the possible effects of fixed pattern noise) assuming that such fixed pattern noise does not preferentially generate features akin to a \neviii\ doublet.

\subsection{Optimised selection of weak  \neviii\ absorption}
\label{weaksearch}

In this section, we describe how we select \neviii\ absorbers in a column-density regime that lies below the S/N-imposed limit for direct line searches. This limit occurs around a column density of log N$\sim$ 13.7, and 
Fig.~\ref{logN_vs_minflux} shows that for such lines  the minimum flux at line centre is above F=0.88. Hence, our first selection criterion is requiring that a pixel has a minimum flux $ F_{\rm {min} } \geq$ 0.88.

\begin{figure}
\includegraphics[angle=270, width=90mm]{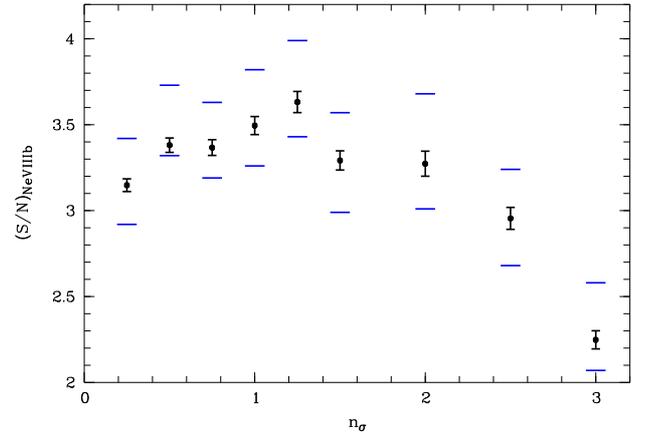}\caption{The procedure to determine the low significance cut-off for the selection of absorbers for one combination of slope and normalisation. We produce a set of 10 different realisations of mock data, and implement a lower flux cut and significance inclusion cut n$_\sigma$. The black data points indicate the mean (and its error) of the expected signal strength for \neviiib, the spread is represented by the blue, vertical lines. Being stricter about the inclusion of pixels leads to a purer sample, yet results in lower S/N for the composite. The sweet spot for the maximisation of the significance of the signal is always close to n$_\sigma \sim$1.0. This example uses the same \neviii\ model as in Fig. \ref{example_composite}, and here we select n$_\sigma$=1.25 as optimal search parameter.} 
\label{optimal_selection}
\end{figure}

On the low flux decrement side, we want to be as inclusive as possible, trying not to exclude the weakest absorbers, which are presumably the most abundant. However, it is clear that, by selecting pixels closer to the 100\% transmission, we include a higher fraction of pixels that are simply pushed to lower than unity flux by noise, thereby increasing signal dilution. We use the following criterion for determining whether pixels at the weak absorption limit be retained
\begin{equation}\label{equation_sigma}
F_{\rm max}\leq 1 - n_{\sigma} \sigma(F) 
\end{equation}
where $\sigma$(F) is the error estimate on the flux F, and n$_{\sigma}$ the minimum number of standard deviations required for absorption identification. In the following, we optimise our choice of  n$_{\sigma}$ backed on mock data to maximise \neviiib\ signal significance.

We quantify \neviiib\ detection significance by comparing the expected strength of the \neviiib\ signal in the composite with the noise of the composite spectrum. We define the potential \neviiib\ S/N  as
\begin{equation}\label{equation_significance}
(S/N)_{\neviiib} = (1 - F_{s, \neviii} )/ \sigma_{s, \neviiib},
\end{equation}
where  $F_{s, \neviiib}$ and $\sigma_{s, \neviiib}$ are the flux and the error estimate of the flux respectively at the pixel centred on \neviiib\ in the composite spectrum. Note that in the following analysis we measure $F_{s, \neviiib}$ both in the real composite and the composite from mock data, but we always use the error estimate of the composite with {\it real} data. The errors in the composite spectra are estimated following the method set out in Appendix~\ref{errors}.

With those definitions in hand, we analyse a set of model spectra to determine the optimal n$_\sigma${} for {\it each} test \neviii\ absorber population parametrised as a CDDF with chosen slope and normalisation. A high n$_\sigma${} means purer selection of absorption, but also reduces the number of pixels selected, resulting in a lower S/N for the stack. At low choice of n$_\sigma${} improves the S/N for the composite as more spectra are stacked, but mixes in more noise. There is an optimal choice for n$_\sigma${} balancing those two effects, as can be seen in Fig. \ref{optimal_selection}. For every test slope and normalisation, we determine the optimal n$_\sigma${} in the following way. We try a range of n$_\sigma$ and flux ranges $0.88 <F \leq 1 - n_{\sigma} \sigma(F) $. For each we calculate the composite spectrum using an unweighted arithmetic mean, and this provides the error estimate $\sigma_{s, \neviiib}$ for each choice of n$_\sigma$. Then for every choice of n$_\sigma$ we compute 10 realisations of the model spectra. For each realisation we select pixels in flux in the range $0.88 <F \leq 1 - n_{\sigma} \sigma(F) $, the composite spectrum is computed and a S/N is projected using Equation~\ref{equation_significance}. The resulting projected S/N for \neviiib\  are averaged over those 10 realisations.

\begin{figure*}
\includegraphics[angle=270,width=140mm]{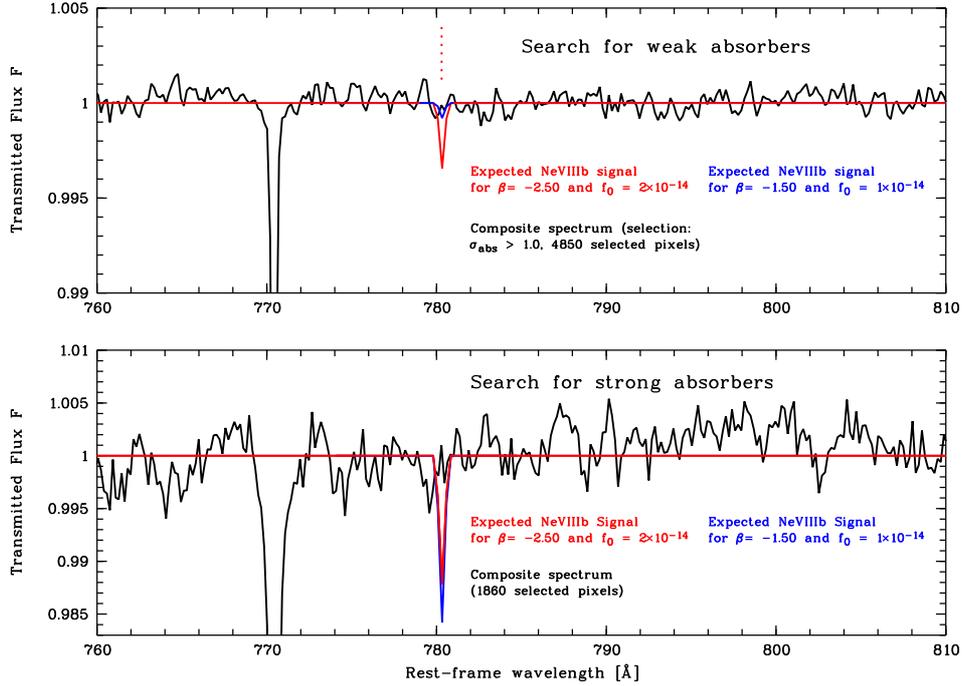}
\caption{Composite spectra in the absorber restframe for two specific assumptions about the absorber distribution in the weak (upper panel) and strong (lower panel) search regime. Note the scale for flux: while the S/N/pixel reaches values $> 1000$ for the upper panel, the strong search results in lower S/N/pixel due to the selection of fewer pixels and a larger flux spread allowed to enter the composite. Shown in red and blue are the expected signals at the location of the weaker doublet member \neviiib, derived by averaging over sets of model spectra. The statistical error for the models can be made arbitrarily low by increasing the number of mock data sets, and hence we have set it to zero here for plotting purposes only. The specific model case for the red CDDF is the same as in the preceding figure. In both search regimes the resulting composite yields a null detection of \neviii\ at a high significance for the CDDF with $\beta =-2.5$, whereas the shallower slope of $\beta=-1.5$ leads to a signal for the weak absorber search much too weak to be detected reliably with our data.}
\label{example_composite}
\end{figure*}

The black data points in Fig. \ref{optimal_selection} represent the S/N derived from mean of these realisations and its error. The blue horizontal lines indicate the range of values found within all 10 realisations. While 10 realisations are low for an accurate estimate of the S/N errors, they are sufficient for a broad optimisation for the wide range of trial CDDFs studied here. Once we select a n$_{\sigma}$ which maximises the S/N of the composite, we rerun the mock analysis with a higher number of realisations in order to reduce the sampling error to below a factor of one half of the error expected from the real data. This typically requires on the order of 500 realisations for each case.
For all the CDDFs parametrisations considered, the optimal n$\sigma${} is always close to a value of 1.0. This is both re-assuring in that our stated goal is to go below column densities required for detection in traditional line searches (those rely upon much higher feature significance), but also that we remain in the realm of at least moderately secure statistical detection.

 \subsection{The selection of strong \neviii\ absorption}
\label{strongsearch}

We can extend our stacking analysis also to the regime of absorbers stronger than log N$\sim$13.7. Although it is less
informative about properties of individual absorbers compared to
direct line searches, it nonetheless allows for robust inferences on global properties. Our choice of treating these two regimes separately is not born out of any expectation that they are physically different, but rather  we chose to perform distinct analyses on empirical grounds. While the weak absorber search above is designed to focus exclusively on absorption that is inaccessible to direct line searches (using current data), the strong \neviii\ search presented here takes an alternative approach that does not seek to discover new \neviii\ absorption, but rather re-assess the CDDF constraints they provide. In this respect we take advantage of our fully blind approach that automatically provides constraints on the potential CDDF.

The selection criteria for strong absorbers is simpler than that for the weak population. We require a pixel to be selected to have flux within a window: 0.75 $< F <$0.88. 
The upper limit coincides with the lower flux limit for the weak absorbers, and the lower limit is obtained by extrapolating Fig. \ref{logN_vs_minflux} to the strongest absorbers expected in practise in a dataset of this size for viable CDDFs. It also allows the selection of every \neviii\ found by direct line searches.

\subsection{Construction of the `agnostic' stack}
\label{construction}

Once we have determined an optimal selection procedure for each assumption of the underlying absorber distribution, we may begin to compare them to the observed agnostic stack.

We construct composite spectra following this procedure: first, we identify all
pixels in our dataset satisfying our simple selection criteria {\it regardless} of our knowledge of their nature as absorbers (hence `agnostic'). We then treat each of these as if they were the strong member of
the \neviii\ 770/780\AA\ doublet, and shift the COS spectrum into the absorber restframe. Note that our choice of rebinning into pixels equidistant in log wavelength-space ensures that all of those shifted spectra retain the correct doublet structure within a full pixel. We generate the arithmetic mean of all these absorber spectra, and arrive at a \neviii\ composite spectrum.  We compute the arithmetic mean, however, we exclude all pixels from the calculation that exhibit a flux too low to be caused by \neviiib. 

For the weak absorber composite, and given the overall noise characteristics of the full dataset, we find that there is a negligible probability that a measured flux $F<0.85$ could  correspond to significant \neviiib\ absorption for such weak \neviiib, even in the presence of plausible noise (i.e. both underestimates of selected \neviiia\ or overestimates of \neviiib\ ). Where the potential \neviiib\ flux is $F<0.85$ we discard the data before stacking. A similar argument holds for the minimum flux in the strong \neviii\ absorber search, which results in a $F>0.75$ limit selection due to this optimsation. 

\begin{figure}
\includegraphics[width=90mm]{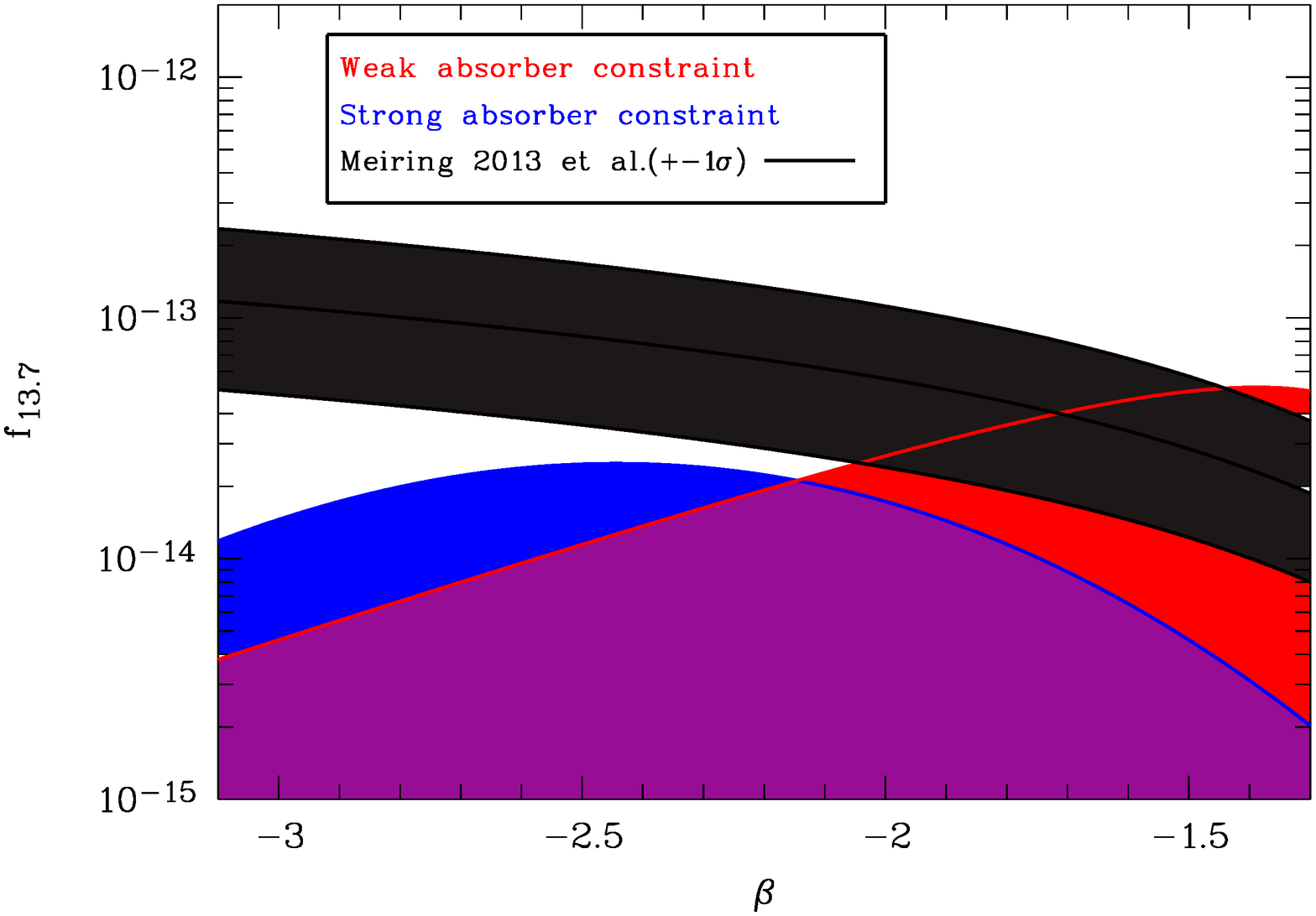}\caption{Constraints on the \neviii population derived independently from the stacking of weak and strong absorption. The model is a single power-law distribution with slope $\beta$, and normalisation f$_{13.7}$, here fixed at log N$ _{\textrm{\neviii}}=13.7$. This model population extends over 12.0 $<$ log $ _{\textrm{\neviii}} <$ 17.0. The red, blue, and purple coloured regions depict areas that are allowed by either (or both) the search for weak (log N$ _{\textrm{\neviii}} \leq 13.7$) or strong absorbers (log N$ _{\textrm{\neviii}} > 13.7$) at least at a 1.0$\sigma${} significance. 
The black band represents the 1$\sigma$ allowed range for a hypothetical absorber population that reproduces the $d\mathcal{N}/dz$  inferred by \citet{Meiring2013}{} for strong absorbers (logN$ _{\textrm{\neviii}} > 13.7$) in PG1148+549.}
\label{separate_constraint}
\end{figure}

Fig. \ref{example_composite}{} 
shows examples of those composites (black curves) together with the \neviiib\ signal expected from the modelling (red and blue curves) for both weak and strong \neviii\ searches. Note that while we created one such composite for each of the assumed CDDFs for the weak \neviii\ absorber distribution, the differences between those composites are minimal because the only free parameter for the selection (n$_\sigma$) varies  very slightly. Hence, the composite for the weak absorber search shown here is illustrative of the full range of composites considered. Note that there is only one composite spectrum for the strong absorber regime since it lacks the adaptive treatment of absorption close to the continuum performed for the weak absorber analysis. This composite spectrum is depicted in the figure in black in the bottom panel. 

The same selection and stacking procedure is applied to the mock data sets. The expected signal strength for the mock absorbers can be estimated directly by measuring the absorption at the position of \neviiib\ in those stacks.

\section{Results}\label{Results}

As detailed in the preceding section, Fig. \ref{example_composite} shows an optimised search for a particular \neviii\ population and shows no detected \neviii\ signal. Indeed, we arrive at a {\it non-detection}{} for any  \neviii\ population in the full range of slopes for the CDDF that we tested (-3.6 $< \beta \leq$-1.3), despite unprecedented sensitivity. In this section we analyse how such a non-detection constrains the \neviii\ population by comparing it with expectations from our mock data for parametrisations of the column density distribution function.

\subsection{Placing limits on the column density distribution function}\label{limits_on_CDDF}

We investigate a potential \neviii\ signal from both weak and strong \neviii\ absorbers, the selection of which is set out in sections \ref{weaksearch} and \ref{strongsearch} respectively. The weak search goes beyond what is possible in direct detection studies by going down to the noise limit to  statistically explore absorbers, which may be individually insignificant. The strong search reassesses the regime that is accessible to direct detection studies, but uses our blind statistical search methodology and the in-built inference of CDDF limits for our full spectroscopic sample. The boundary between these two searches is at a flux of $F=0.88$, which corresponds to a column density of log $N_{\textrm{\neviii}} =  13.7$.

We  explore allowed values for the CDDF parameters $\beta$ and $f_{13.7}$. We  treat the weak search and the strong search as separate regimes divided at log  $N_{\textrm{\neviii}} =  13.7$, but also combine the constraints across the full  accessible range of columns. It should be noted, however, that constraints in two column density bands should not be interpreted as a detection of a broken power-law fit with a break at log  $N_{\textrm{\neviii}} =  13.7$. Nor should a single power-law constraint be viewed as an indication that a single power-law fit is favourable. Both choices are simply acceptable ways of expressing our constraints.

Figure~\ref{separate_constraint}{} shows constraints in $\beta$ and $f_{13.7}${} derived from our weak and strong searches treated independently. The red region indicates the one standard deviation allowed range for our weak absorber search and the blue contour shows our one standard deviation allowed range for our strong absorber search. CDDFs allowed by both are shown in purple. Also shown (in black) are the CDDFs that are consistent with the \citet{Meiring2013} 1$\sigma$ uncertainty in $dn /dz$ at log $ N_{\textrm{\neviii}}  \approx 14.3$.  Our allowed ranges in $\beta$ and $f$ for the strong absorbers are in tension with the the \citet{Meiring2013} $dn / dz$ at a 2.2$\sigma$ level. We fit the 1$\sigma$ contour for the weak absorber search with a 3rd order polynomial in $\beta$, and obtain for the normalisation 
\begin{equation}
\log f_{13.7} < -14.824 -2.6256 \times \beta -1.3484 \times \beta ^{2} -0.17559 \times \beta ^{3}. 
\end{equation}
The same procedure for the  1$\sigma$ contour for the strong absorber search yields 
\begin{equation}
\log f_{13.7} < -18.933 - 4.6383 \times \beta -1.1200 \times \beta ^{2} -0.04668 \times \beta ^{3}.
\end{equation}

Note that the strong \neviii\ absorption constraint in Fig. \ref{separate_constraint}{} implies a maximum allowed CDDF amplitude $f_{13.7}$ for any chosen slope $\beta$. This maximum $f_{13.7} = 2 \times 10^{-14}$ occurs at $\beta=-2.44$.

\begin{figure*}
\includegraphics[width=120mm]{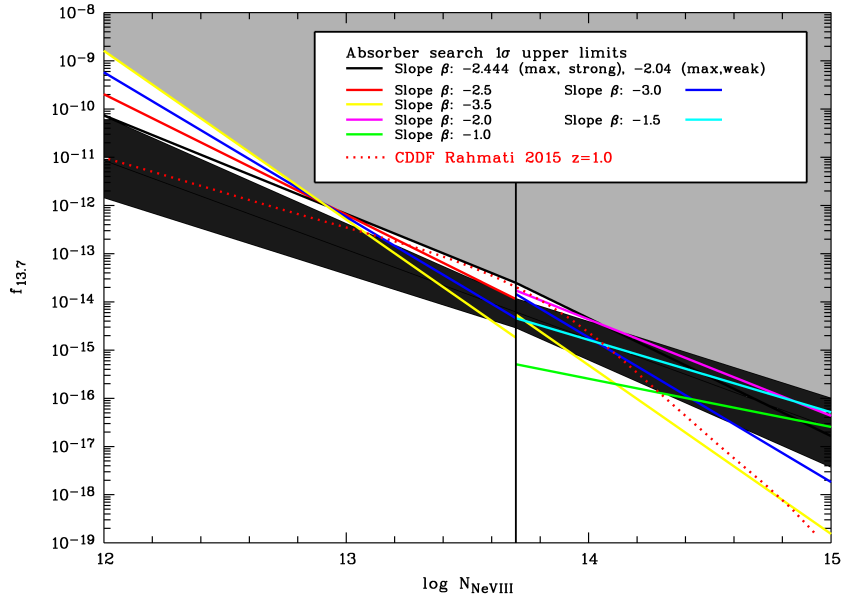}\caption{1$\sigma${} upper limits on the normalisations of the CDDFs under different scenarios for the slope of the assumed power-law distribution of the \neviii\ population. The solid lines beyond log N$ _{\textrm{\neviii}} >$13.7 are derived by the search for such strong absorbers, whereas the solid lines for log N$ _{\textrm{\neviii}}  \leq 13.7$ stem from our search for the weak population. The area in gray marks parameter space that is ruled out regardless of the CDDF slope at least at the 1$\sigma${} level. The normalisation of the CDDF is quoted at the boundary between the weak and strong regimes.  As Figure 6 shows the maximum allowed normalisation for the weak search is higher than that of the strong, so for continuity reasons we limit the weak regime CDDFs shown to those allowed by the strong search. The dotted line represents the  CDDF derived by simulations at similar redshifts \citet{Rahmati2015}(which we test in Section \ref{Discussion}). The area in black delineates our best estimate for the CDDF based upon combining the stacking with the direct detections.}
\label{different_CDDFs}
\end{figure*}

In Figure~\ref{different_CDDFs}, we show 1$\sigma$ upper limits on the column density distribution function normalisation $f_{13.7}$ (for a range of values of $\beta$) for both our weak \neviii\ search  and our strong \neviii\ search. Also shown for reference is the simulated CDDF from  \citep{Rahmati2015}.
The area in gray is ruled out by all possible choices for the slope in the two different regimes at the 1$\sigma${} level. 
We constrain the normalisation of both the weak and the strong \neviii\ populations at their interface, log  $N_{\textrm{\neviii}} =  13.7$. A convenient consequence of this choice is that the strong search implies a clear upper limit to this interface incidence rate displayed as a maximum $f_{13.7}$ shown in Figure~\ref{separate_constraint}{} (at  $\beta=-2.44$). This also sets a continuity requirement to our weak absorption analysis. As such only parametrisations of this weak sample that do not break this log  $N_{\textrm{\neviii}} =  13.7$ continuity requirement are considered in Figure~\ref{different_CDDFs}. This figure shows all power-law slopes that constrain the 1$\sigma$ allowed CDDF with additional slopes show for illustration. One can take this continuity argument a step further: for any chosen  normalisation $f_{13.7}$ one may place a horizontal line on Figure~\ref{separate_constraint}{} and read off all the allowed slopes (at the 1$\sigma$ level) for both the weak and strong populations.

\begin{figure}

\includegraphics[width=90mm]{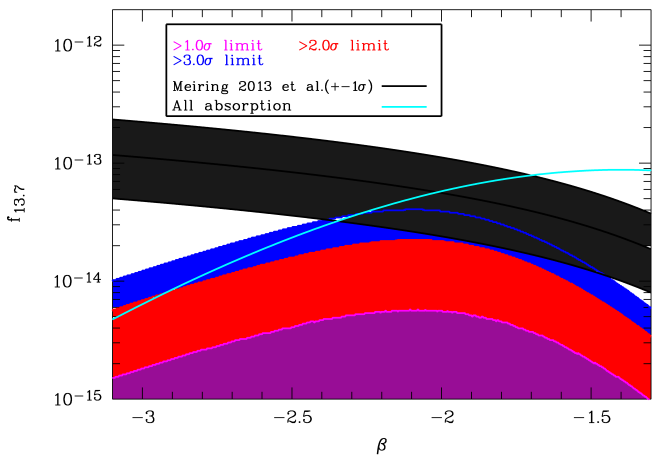}\caption{Limits on the \neviii\ population when combining the results of the searches for weak and strong absorption. Assuming a single-power distribution for the \neviii\ over the complete range of column densities probed by us (12.3 $<$ log N$ _{\textrm{\neviii}} <$ 15.0), we can combine both independent approaches to obtain tighter constraints than the individual regimes allow. The areas coloured in purple, red and blue delineate the boundaries for parameter space allowed by this combined analysis at the $>1.0, >2.0$, and $>3.0\sigma$ level. In the main text, we give the coefficients for a 3rd-order polynomial fit to those curves. The black band is the same as in Fig. \ref{separate_constraint}. The cyan line represents an upper limit derived by comparing single-power distributions with the entire population of apparent absorption using our mocks as set out by Appendix~\ref{limitsfrommocks}.}
\label{combined_constraint}
\end{figure}

In addition to the above independent assessment of the weak and strong \neviii\ regimes we also  explore constraints assuming a single power-law CDDF combining both searches. The allowed values of slope and amplitude for such a power-law are shown in Figure~\ref{combined_constraint} for one, two and three standard deviations. 
These combined limits are our main result, and constitute the tightest constraints thus far on the intervening \neviii\  absorber population at the redshifts considered here over the range of column densities 12.3 $\leq$log N$\leq$15.0. Fitting the 1$\sigma$ allowed contour boundary with a 3rd order polynomial in $\beta$, we obtain for the normalisation of the CDDF 
\begin{equation}
 \log f_{13.7}  < -21.46 -8.366 \times \beta -3.068 \times \beta^{2} -0.342 \times \beta^{3}. 
 \end{equation}
 Also over-plotted is the range of allowed CDDF parameters inferred from the observed $dn /dz$ at log $ N_{\textrm{\neviii}} \approx 14.3${} from \citet{Meiring2013}.
The cyan line in Fig. \ref{combined_constraint} represents an upper limit to the absorber population that is derived without stacking the data at all, but by analysing 
limits on the expected combined population of other absorbers and the scope for residual \neviii\ in the measured flux PDF.
The details of this procedure are summarised in the Appendix~\ref{errors}.

\subsection{The inferred mass density of \neviii}
The cosmological mass density in units of the critical density, $\Omega = \rho / \rho _c$, traced by a particular ion can be estimated by 
\begin{equation}\label{ion_density}
\Omega _{ion} = \frac{m_{ion}}{\rho _{c}}(\frac{c}{H_{0}}\Delta X  _{tot})^{-1}N^{tot}_{ion}
\end{equation}
where m$_{ion}${} is the atomic mass, $\rho_{c}${} the critical cosmic density, $\Delta X _{tot}$ the total absorption distance in the survey, and $N^{tot}_{ion}$ the total column density of all absorbers.

We obtain upper limits to the total surveyed \neviii\ mass density by integrating over the CDDFs underlying our models. We again treat the two column density ranges for weak (12.3$<$ log N$ _{\textrm{\neviii}}  \leq 13.7$) and strong absorbers (13.7$<$ log N$ _{\textrm{\neviii}} <$15.0) separately, because both search methods are tailored to yield the tightest constraints in their respective column density regime.

\begin{figure}
\includegraphics[angle=270,width=90mm]{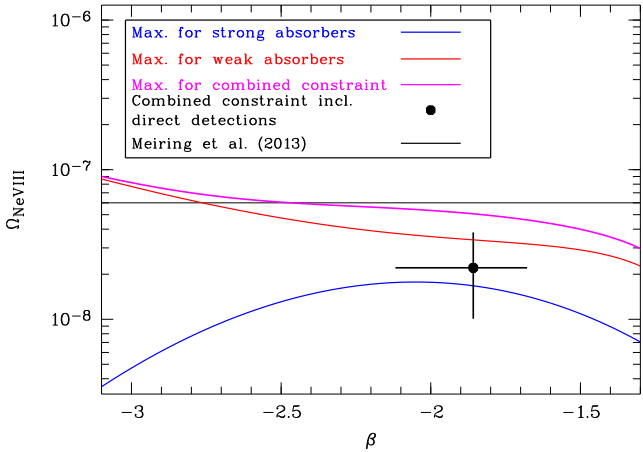}\caption{The 1$\sigma$ upper limit for the density of the \neviii\ absorber populations probed by our two search methods. The red line indicates the upper limit for absorbers falling into the regime 12.3 $< \textrm{log} N_{\textrm{\neviii}} < 13.7$. The black line is the upper limit for stronger absorbers (13.7 $< \textrm{log} N_{\textrm{\neviii}} < 15.0$). The upper limit of our strong absorber bin is set by the observational fact that none of the detected \neviii\ absorbers have column densities log N$\geq 14.7$, whereas the lower limit marks the boundary where weaker absorbers stop to have an impact on the expected signal. The magenta line is the result of the combined constraint. The horizontal line in black is the density estimate by \citet{Meiring2013}, whose 3 detected absorbers above log N=13.7 do not allow for a constraint on the slope of the CDDF. The estimate for the black data-point results from the best constraint when combining the stacking with data from direct detections.}  
\label{density_limits}
\end{figure}

Figure \ref{density_limits}{} shows the 1.0$\sigma${} upper limits to the density derived by our searches. Also shown is a density derived from the \citet{Meiring2013} number of systems per unit redshift ($d\mathcal{N}/dz = 7^{+7}_{-3}$) in the spectrum of PG1148+549 (see Section~\ref{PG1148+549} for further discussion of this case). Note that this $d\mathcal{N}/dz$ is directly comparable to our upper limit for the strong search only since this probes an equivalent range of column densities. 

For slopes $\beta < -2.0$, the density budget becomes dominated by the highest column density systems, and hence in those cases the upper integration limit is of crucial importance when comparing between different density estimates. None of the detected intervening \neviii\ absorbers have log N$\geq$14.7, and hence our range extending up to log N=15.0 is well chosen for a comparison. Likewise, for slopes $\beta <-2.0$ the lower integration limit plays the dominant role in setting the density. The red dashed line in Figure \ref{density_limits}{} represents the density contained in absorbers of column densities $ 12.3 \leq$N$ _{\textrm{\neviii}}\leq$13.7. The lower limit here is chosen such that we exclude very weak absorbers that do not contribute to the \neviiib\ signal, even if selected.

If we combine both searches by demanding the CDDF to be represented by one single power-law over the whole range of column densities, we obtain the limit delineated by the magenta solid line in Figure \ref{density_limits}. While this combined mass density is superficially similar to the value derived from  \citet{Meiring2013} this is purely coincidental as we integrate over a column density range nearly two orders of magnitude wider. As we shall see in Section~\ref{combinedres} this constraint on the cosmic \neviii\ density can be turned into a measurement when combined with a conservative assessment of the directly detected \neviii\ population.

\begin{table*}
\caption{Basic properties of all directly detected \neviii\ absorber components that exhibit a column density log N$_{\neviiia} \ge 13.7$, i.e. our strong absorber criterion. Note that the estimate for the column density depends on the measurement method, in general, column densities derived via the apparent optical depth method (AOD) are higher than those for Voigt profile (VP) fits of the same feature. The temperature and metallicity estimates quoted here are the results of CLOUDY modelling for a variety of low ionisation and high ionisation transitions in those absorbers. Ionization and metallicity calculations assuming collisional ionization are indicated by `CIE' and photoionization by PIE.}
\label{tab:direct_detections}
\begin{tabular}{@{}lccccccc}
  Sightline        & z$_{abs}$ & log N$\_\neviii$ & b [km/s] & Metallicity [Z/H] & Temperature [K] & Reference  \\
\hline

pks0405   & 0.4951 & 13.96$\pm$0.06 & 70$\pm$4  & -0.35 (CIE), -0.6$\pm$0.3 (hybrid) & 5$\times 10^5$ & \citet{Narayanan2011} \\
s080908   & 0.61907 & 13.76$\pm$0.14(AOD) & 69$\pm$20 & $<$0.5 & $>$4$\times 10^5$(CIE) & \citet{Pachat2017} \\
sbs1122   & 0.57052 & 13.72$\pm$0.15(AOD) & 38.8$\pm$8 & $<$-0.8 & $10^5$ &  \\
pg1407    & 0.5996 & 13.80$\pm$0.15(VP) & 31$\pm$15 & $>$0.0 (PIE) & 6.3$\times 10^4$ (PIE) & \citet{Hussain2015} \\
          &        &                               &           &  $>$-1.0 (CIE) & 5$\times 10^5$(CIE) &  & \\
pg1148    & 0.68381  & 13.98$\pm$0.09 & 32$\pm$5 & $>$-0.5 & 5$\times 10^5$ &  \citet{Meiring2013} \\
          & 0.70152  & 13.75$\pm$0.07 & 28.2$\pm$7.1 & $>$-0.5 & 5$\times 10^5$ & \\
          & 0.72478  & 13.70$\pm$0.12 & 41.4$\pm$7.5 & $>$-0.5 & 5$\times 10^5$ & \\

pg1206    & 0.927 & 13.71$\pm$0.29 &  & 0.48 (from local gas) & 2.2-3.8$\times 10^5$ & \citet{Tripp2011} \\
          & 0.927 & 14.04$\pm$0.08 &  & 0.0 (from local gas) & 3.3-4.0$\times 10^5$  &  \\
          & 0.927 & 14.07$\pm$0.04 &  & - & - &  \\
          & 0.927 & 14.53$\pm$0.04 &  & -0.3 & 4.2-4.5$\times 10^5$ & \\
          & 0.927 & 14.21$\pm$0.05 &  & - & - &  \\
          & 0.927 & 13.78$\pm$0.09 &  & 0.0 & - & \\
\hline

\end{tabular}
\end{table*}

\subsection{Combination with direct line search results}
\label{combineintro}

Here we investigate the viability of a particular \neviii\ population (as characterised by its CDDF) in light of current statistics of individual  identifications of \neviii\ absorbers found in our sample in combination with our agnostic stacking results. We do this by initially limiting our sample to those of sufficient median S/N such that all identified \neviii\ systems can be reliably identified. Then we analyse the likelihood that a particular CDDF is in agreement with the observational constraints in a two step process. In the first step we determine the fraction of realisations of our mock spectra suite to generate the desired direct line detection properties (see below). This provides a conservative likelihood that the trial CDDF is consistent with the desired direct line statistics. This is not to say that  these strong metals lines would be detectable in a blind analysis of the mock spectra (they may, for example, be excessively blended with contaminating absorption); it is a minimal statement that, for the number of lines required, there must be at least this many {\it present} in the perfect mock data.

In the second step we retain the mock realisations with a sufficient number of directly detectable lines, and pass only mock suites `surviving' through the analysis with our agnostic stacking as described in Section \ref{limits_on_CDDF}. In fact, we perform both the weak and strong \neviii\ agnostic stacking analysis and combined their likelihoods to a single likelihood in this case. The product of the two fractional probabilities from these two steps, constitutes the combined likelihood (or joint probability) that a model CDDF is consistent with both the direct line search statistic desired and our agnostic stacking.

We proceed to consider two characterisations of  direct line detections in our data set. Initially we ask whether the direct detections of 3 \neviii\ lines seen in quasar PG1148+549 is consistent with our implicit assumption that observed \neviii\ systems are physically distinct and independent from one another. We then ask what single power-law models are consistent with both our agnostic stacking and the incidence rate of directly detected \neviii\ lines overall. 

\subsubsection{The atypical case of PG1148+549}
\label{PG1148+549}

Along the line of sight towards PG1148+549+549 at $z_{em} = 0.9759$ (part of our analysis sample, see Table~\ref{tab:sightlines}), one of the spectra with the highest overall S/N, there are not just one, but three individual, strong (log N$ _{\textrm{\neviii}} >$13.7) \neviii\ absorber systems at redshifts z$_{abs} =$0.68381, 0.70152, and 0.72478. This separation of $\delta z=0.04$ constitutes $7200 km/s$ if treated as a peculiar motion, or a comoving distance of 83 h$^{-1}$Mpc if treated as part of the Hubble Flow. 
For details of these detections see \citep{Meiring2013}. Limiting oneself to the portion of our sample which has sufficiently high signal-to-noise to confidently detect these systems (S/N $>$20 per pixel), one retains most of our analysis path ($\delta z=0.43$) for this spectrum. Also the redshifts are typical of our sample, so based on these most simple criteria PG1148+549 is unexceptional. However, since all other sightlines of such quality show only one or no such absorbers (and only 5 in total), it seems that the sightline towards PG1148+549 is atypical.
We quantify the likelihood that this state of affairs arises by chance and in light of our data analysis of 26 sightlines.

The high value for $d\mathcal{N}/dz$ one would derive on the basis of this one sightline alone can be ruled out as a good measure of the {\it cosmological} $d\mathcal{N}/dz$ with confidence. 
Here we go a step further and assess the joint probability of our agnostic stacking and these direct line detections. We do this by using the two step process outlined in general terms in Section~\ref{combineintro} above.

We limit ourselves to only those spectra in our sample that exhibit a median S/N that is comparable to the PG1148+549 spectrum and so allow for the direct detection of \neviii\ absorbers of the column density range sampled by \citep{Meiring2013}. This reduces the original path to $dz=4.73$ or 11400 (rebinned) pixels. We randomly populate those pixels with absorbers drawn from a trial CDDF. Then we divide the total path into equally-sized portions of the size used by \citep{Meiring2013}. We determine what fraction of those mock realisations resemble the real data by requiring that one portion must have at least three strong absorbers, while all other ones must have either none or one. Furthermore, the redshift separation of the 3 absorbers in that one special bin must resemble the spread seen in the detection of \citet{Meiring2013}. This constitutes step one and provides the probability that the model CDDF is consistent with having an outlier quasar sightline of this type. In the second step, we further assess the mocks which meet the previous criterion for consistency with both our weak and strong agnostic stacking procedures.

\begin{figure}
\includegraphics[width=90mm]{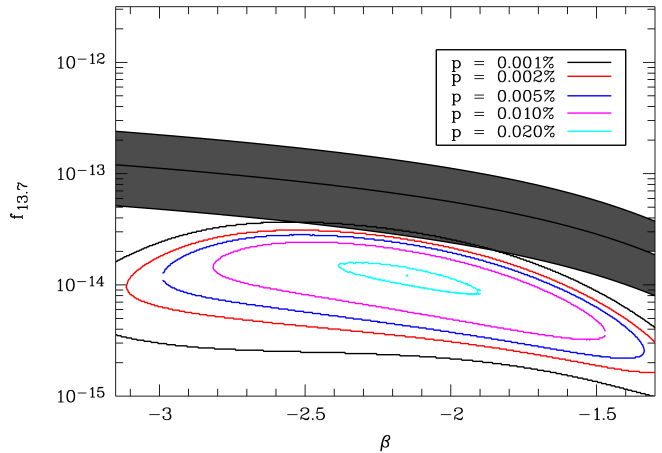}\caption{The joint probability of our combined stacking constraint and having three strong absorber systems (as would appear to be the case for PG1148+549) in one spectrum occurring by chance and no more than one in every other spectrum. Even in the most favourable case, at a normalisation an order of magnitude lower than indicated by the $d\mathcal{N}/dz$ of \citep{Meiring2013} (black band) is required. This probability peaks only at 0.024$\%$, indicating that at least two of these lines are very likely to be associated with a single system.}  
\label{meiring_test}
\end{figure}

The result shown Figure \ref{meiring_test} is a joint probability of our stacking measurements and the direct line detections of \neviii\ in PG1148+549 assuming a single power-law CDDF. This figure shows the 1$\sigma$ allowed range of CDDF that one derives from the $d\mathcal{N}/dz$ of \citet{Meiring2013} as a black band. We find that the optimal CDDF has a likelhood of only 0.024\%. This is (as \citealt{Meiring2013} points out) confirmation that the sightline is probing a highly unusual large-scale structure, that should be further scrutinized. 

\subsubsection{Combined constraints with the overall incidence rate of direct detections}
\label{combinedres}

We now combine constraints from both our agnostic stacking technique and direct line detection studies. We limit ourselves to data of sufficient S/N for direct line studies (as in the case for PG1148+549, we require S/N per pixel $\geq$ 20), and we further limit ourselves only to the direct detections found in our sample path. A detailed analysis of all the absorber systems reveals that there are 13 such individual components that satisfy the following two criteria to be included in our statistical analysis: a column density log N$ _{\textrm{\neviii}} >$13.7, and an individual velocity structure such that the separation from other components is large enough for the feature to occupy one distinct pixel, i.e. $\delta v \geq$80 km/s to the centre of the next component. The components that fulfil these criteria are shown in Table~\ref{tab:direct_detections}). The direct detection of some of these components show limitations that cast some doubt on  their identification as \neviii\ (blending of one or both parts of the \neviii\ doublet with unrelated  absorption, lack of accompanying additional transitions such as \ovi\ at the same velocity, or individual detections with less than 3$\sigma$ significance). Hence, in the following we vary the number of assumed direct detections and allow the reader to make his/her own assessment of the number of true \neviii\ direct detections in this sample. With the direct detection components, $n_{comp}\le 13$, we explore the joint probability combining with our agnostic stacking results and assess the resultant constraints on the CDDF. 

\begin{figure}
\includegraphics[width=90mm]{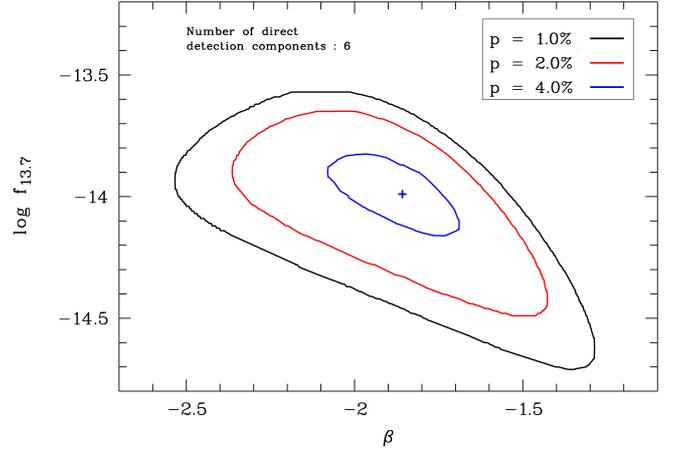}\caption{The joint probability of consistency between a single power-law column density distribution function of normalisation $f_{13.7}${} and slope $\beta$ in light of  observational constraints for our sample of spectra. The direct line constraint is allowed for by requiring at least 6 components sufficiently strong for direct identification. (For more details see text.)}  
\label{combined_direct}
\end{figure}

For the joint probability analysis of stacking and direct detections we follow a variant of the procedure set out in Section~\ref{PG1148+549}. For every allowed value of $n_{comp}$, we assess the fraction of mock suites with $n_{comp}$ components with logN$ _{\textrm{\neviii}} >$13.7 in the entire path with S/N/pixel$\geq$ 20, and treat this as the probability that a sufficient number of strong absorber be present for a particular trial CDDF.  We then use these remaining mock suites to determine the probability of consistency with our agnostic stacking measurements.  The  resultant two-step joint probability for $n_{comp}=6$ is shown in Figure~\ref{combined_direct}. For each choice of $n_{comp}$ we can hence determine the maximum likelihood, and the parameter combination of the CDDF that provides the optimal constraint under that scenario.       

\begin{figure}
\includegraphics[width=90mm]{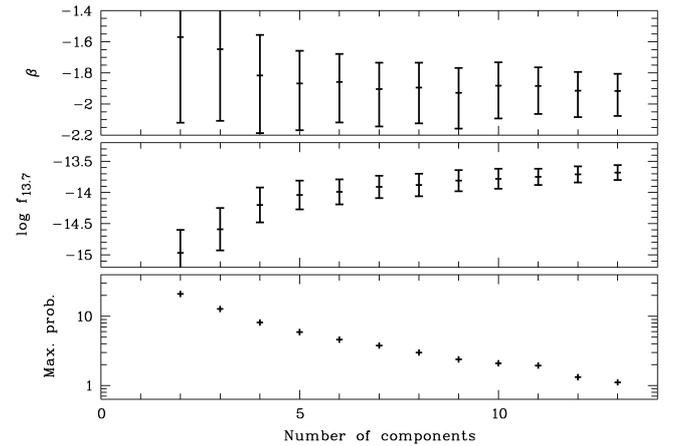}\caption{Lower panel : The maximum joint probability of consistency between a single power-law column density distribution function of normalisation $f_{13.7}${} and slope $\beta$ as a function of requiring a number $n_{comp}$ of components sufficiently strong for direct identification. This maximum likelihood is reached in each case by the CDDF with normalisation $f_{13.7}${} and slope $\beta$ indicated in the middle and upper panel.}  
\label{max_prob_different_n}
\end{figure}

Figure~\ref{max_prob_different_n}{} shows how the maximum likelihood varies with $n_{comp}$ (lower panel), and how the different choices for $n_{comp}$ affect the optimal parameters of the CDDF (middle and upper panel). It is evident that demanding a higher number $n_{comp}$ of strong absorber components to be present in the line list while maintaining a null detection during the stacking analysis becomes increasingly improbable (despite the small numbers considered), and hence there is clear statistical tension between our stacking results and larger allowed direct detection populations ( $\sigma \approx 2.5$ for $n_{comp}=13$). The CDDF parameter combinations, however, which pick out this maximum likelihood in each case are within  1$\sigma${} of each other, once the number of components $n_{comp}>8$.  Specifically, the $\beta$ of the CDDF is well converged, and the rise in the normalisation log f$_{13.7}$ from $n_{comp}=8$ up to maximum, $n_{comp}=13$, is only marginal. 

Our modelling does not take into account complex velocity structures of components extended over sometimes high velocities (see the extreme case described in Section~\ref{PG1148+549}), which acts to boost the number of distinct components observed to greater than the number of source systems. Hence we favour a refined sample of direct detection components with only 1 component counted within 1500 $km/s$. This results in a favoured number $n_{comp}=6$, (reflecting additionally the merging of 3 components to 1 absorber in the case of PG1148+549, along with similar complexes as indicated in Table~\ref{tab:direct_detections}). 
For $n_{comp}=6$, the best fitting CDDF corresponds to $\beta = -1.86 \substack{+0.18 \\ -0.26}$ and log $f_{13.7} = -13.99 \substack{+0.20\\-0.23}$. The pathlength density of \neviii\ absorbers given by theses CDDFs for $n_{comp}=6$ is $dn/dz(z=0.88)=1.38 \substack{+0.97 \\ -0.82}$.
Armed with a CDDF constraint and assuming a single power-law, we can now provide a measurement of the cosmic density of \neviii\ of 
\begin{equation}
\Omega  _{\textrm{\neviii}} (12.3 \leq \log N \leq 15.0) = 2.2 \substack{+1.6\\ -1.2} \times 10^{-8}.
\end{equation}
The upper integration limit for the column density (log N$ _{\textrm{\neviii}} \leq 15.0${} can be justified by the observational result that no detected absorbers have be found with higher column densities, the lower limit reflects the detection limit of our agnostic stacking method.

The favoured single power-law CDDF in our analysis is given by a slope of $\beta = -1.86$, which leads to significant \neviii\ mass in both the strong absorber regime and the weak absorber regime, but the dominant contribution is in the strong absorber regime. Specifically 59\% of the \neviii\ mass is found to reside in the interval $13.7 < {\rm log} N \leq 15$, while we infer that 41\% resides in the weak absorber regime $12.3 < {\rm log} N \leq 13.7$.

Given this \neviii\ density we can infer the cosmic baryon density of the \neviii-bearing gas. Since the dominant contribution to the \neviii\ mass arises in the strong absorber regime of direct detections, we take conditions typical for direct detections in Table~\ref{tab:direct_detections}: a solar metallicity, ionisation fraction  $f(\neviii /\textrm {Ne})=0.13$ \citep{Gnat2007} reflecting a typical temperature of  $T=5\times 10^{5}$K . Assuming a solar neon relative abundance of log$_{10} (n _{\textrm{\neviii}} / n_{H})_{\odot} = -3.91$ \citep{Anders1989}, and atomic weight $\mu$=1.3 to account for He, we estimate a  baryon mass of the \neviii -bearing gas in our sample of 
\begin{equation}
\Omega _{b} \sim 1.8\times 10^{-3} \times \frac {0.13}{f_{\rm NeVIII /Ne}}\times (10^{[Ne/H]})^{-1}.
\end{equation}
While this constitutes only $\sim 4\%${} of the cosmic baryon budget $\Omega _{b} = 0.04907 \pm 0.0014$ \citep{Planck2016}, we note that the ionisation corrections, abundance and metallicity uncertainties render this lower limit very insecure. Interestingly, \citet{Meiring2013} estimate a similar baryon-density for their \neviii-bearing gas, based upon inferences on the total hydrogen columns sampled by their sightline towards PG1148+549. 

\section{Discussion}\label{Discussion}

In this paper we provide new measurements of \neviii\ absorption at a median (mean) redshift z$_{abs}$=0.63(0.88) through agnostic stacking methods using archival COS spectra and in combination with previous searches for \neviii\ in the same sample of 26 high-quality COS spectra (which found 6 such absorbers). These COS spectra were selected only from programs {\it not} directed specifically towards sight-lines proximate to galaxies. Also they were selected purely on redshift and signal-to-noise ratio; the presence or absence of \neviii\ known detections did not play a role. Hence we construct a fully random representative sample of potential \neviii\ absorbing path.

We probe the population of weak \neviii\ absorbers (12.3 $<$ log N $\leq$ 13.7), and the strong \neviii\ population  (log N$\geq$13.7) separately by stacking the appropriate apparent absorption strength. In order to interpret these measurements we construct a forward modelling framework to test the viability of \neviii\ model CDDFs. In the strong absorber regime (log N$\geq$13.7), direct line detections are possible (though rare), but our methods provides a blind approach that also generates a constraint on the CDDF by construction. It should be noted that our approach does not provide lines to be studied individually  e.g. for their thermal/turbulent broadening properties. In both regimes we find an informative null result for \neviii\ absorption.

This forward modelling framework also allows us to combine our constraints on the CDDF from each probe, and further to fold in known \neviii\ detections. Our statistical analysis allows us to place tight limits on the slope and normalization of the \neviii\ CDDF, and thereby characterise the whole population. This represents an advantage over published direct individual detections of \neviii\ lines, which may be drawn from unrepresentative subsets of the available COS spectra (through incomplete publication of sight-lines without detected \neviii\ or the use of sight-lines especially selected for observation due to proximity to known galaxies). Despite some detections of \neviii\ the community has lacked any meaningful \neviii\ CDDF measurement, due to the large uncertainty associated with small numbers and the complexity of folding in analysis differences and uncertain completeness (from both a potential lack of published non-detections and spectroscopic noise or contamination). We address these limitations by treating known detections as a prior for our analysis that digest the small numbers in question while making no assumptions about additional undetected lines present in the path. Key to this is the combination with our agnostic stacking methods and our forward modelling framework.

We note that the large-scale clustering of \neviii\ evident in our analysis (see Section~\ref{PG1148+549}) is absent from our forward modelling, which has potential consequences for our measurement. Referring back to Fig.~\ref{example_composite}, the position at the postulated wavelength of \neviiib\ (780\AA)  is not a random location in the agnostic stack. It is located $\sim$3900 km/s from the potential \neviiia\ absorber selected. If the stacked absorber is indeed \neviiia, and given that \neviiia\ can be clustered on scales $>$7500 km/s, there is an excess probability of NeVIIIa contamination compared to a location in the spectrum far from any \neviiia. Whatever the functional form of this probability distribution function, it should be, on average, a monotonically declining function with increasing line of sight deviation from the selected excess NeVIIIa. As a result, one need not be concerned about an absorption spike mimicking NeVIIIb, but the PDF of contaminating absorption may be different in these portions of the spectra than that of the ensemble average PDF illustrated in Fig~\ref{normalised_pdf_mock_vs_real}. Since our mock-making method generates contaminating absorption to fit that ensemble average PDF, our mocks may be biased with respect to the measurement in the data.

Fig.~\ref{example_composite} demonstrates that this potential systematic bias is negligible with respect to our stochastic uncertainty. If there were a significant large-scale trend in the \neviiia\ signal reflecting the autocorrelation of \neviiia, it would generate a monotonically increasing flux transmission with respect to deviations from 770 \AA, this is not seen in either stacking regime. Moreover, one can see by eye that the noise fluctuations in the composite spectrum beyond 767-773 \AA\ are larger than the deviation from 1 on all scales. Whatever large-scale NeVIIIa signal is present in our stacking analysis it is negligible on scales beyond $\sim$1300 km/s.

In this investigation we assume a single power-law CDDF in scans of parameter space (and test published complex CDDFs predicted from hydro simulations - see below). More complex forms of the CDDF may be parametrised in the forward modelling but are unconstrained by current measurements.  We are able to measure the cosmic density of \neviii\  in our entire column density range 12.3 $<$ log N $\leq$ 15, assuming a single power-law CDDF. A tension between direct detections and our agnostic stacking method is evident but there is no motivation to appeal to a broken power-law CDDF since the tension is driven by the stacking in the same strong absorber regime (log N $= 13.7$) as the direct detections. It should be noted though that, since we do not consider any detections in weak absorber regime, a flattening of the CDDF slope at log N $< 13.7$ cannot be ruled out and would lead to a lower quoted $\Omega_{\textrm{\neviii}}$. 

Note that additional \neviii\ systems have been found outside of our sample (HE0226-4110, \citet{Savage2005, Savage2011} with FUSE/STIS, and 3C263 at z$_{abs}=0.326$ \citet{Narayanan2009, Narayanan2012}. The incidence rate of \neviii\ absorbers per spectrum in these spectra is consistent with our findings (i.e. one per spectrum), but in absence of the statistics of unpublished spectra without \neviii\ detections we are unable to ascertain whether this additional sample as a whole is consistent. The only known exceptional sightline is PG1148+549 \citep{Meiring2013}. PG1148+549 is in fact part of our sample,  and it shows 3 absorbers in one sightline. We have demonstrated that this spectrum is not consistent with the \neviii\ population in the rest of our sample with 0.024\% confidence.

It is informative to test predicted \neviii\ populations derived from hydro simulations using our analysis. The extremely steep slope found in the models by \citep{TepperGarcia2013}{} ($\beta=-2.9$) in combination with the high observed $d\mathcal{N}/dz$ at higher column density can be ruled out completely on the grounds that such a population of weak \neviii\ absorbers would vastly overproduce the total absorption seen in the data. Thus, either the slope has to be much shallower, or the normalisation smaller by almost two orders of magnitude. In this regard, a comparison with the CDDF of \citet{Rahmati2015} derived from the EAGLE simulations (Fig. \ref{different_CDDFs}) is more encouraging. If we compare the CDDF predicted by \citet{Rahmati2015} for their outputs at redshifts $z=1.0$ and $z=0.0$ with our constraints from stacking only, we find that too much \neviii\ is produced, and the model is disfavoured at 0.8$\sigma$ and 1.3$\sigma$ confidence level respectively. Hence our null result from stacking only is in mild tension with their estimate for the \neviii\ CDDF.

We can compare the \citet{Rahmati2015} \neviii\ prediction with our combined analysis of all \neviii\ (both stacking and direct) detections, generating 2.2$\sigma$ tension for the $z=1.0$ output and 2.7$\sigma$ tension for the $z=0.0$ output. 
This tension directly translates into a similar tension for the best estimate of the cosmological mass density, while the CDDF of \citet{Rahmati2015}{} results in a density of $\Omega  _{\textrm{\neviii}} (z=0.7) \sim 6.6 \times 10^{-8}$ over $12.3 \leq \log N \leq 15.0$ (see their figure 5, adjusted to our different column density limits), our best CDDF constraint leads to $\Omega  _{\textrm{\neviii}} (z=0.7) \sim (2.08 \substack{+1.60\\-1.20}) \times 10^{-8}$ (see Figure~\ref{density_limits}). While the data do not demand sophisticated forms for the CDDF in blind scans of parameter space, it is a trivial matter to re-apply this analysis to future predicted \neviii\ populations.

\citet{Rahmati2015} argue that the broad agreement of their EAGLE simulations with metal measurements indicates that the simulations are giving a realistic metal production rate from stellar feedback.It is, however, noteworthy that these and other  simulations ( e.g. \citealt{Hummels2013, Ford2016}) lack \ovi\ in circumgalactic medium regions compared to observations at $z<1$ \citep{Danforth2008, Thom2008, Tumlinson2011, Shull2014}. Flickering active galactic nuclei boost the \ovi\ mass \citep{Oppenheimer2017}, but this correction is likely to exacerbate the tension with our results by generating additional \neviii, since the boost is not exclusively connected to increased photoionization but also collisional ionization. Under the assumption of unchanged relative abundance between neon and oxygen, a picture may be emerging where temperatures are generically over-estimated, since a decrease in temperature would simultaneously generate less \neviii\ and generate more \ovi. This picture is consistent with our low estimate of the cosmic density of \nevii-bearing gas. We stress that despite this low estimate of the cosmic density of \nevii-bearing gas, there is scope for further WHIM gas at low metallicities and/or at higher and lower temperatures compared to the \neviii\ tracing range ($4\times10^{5} K < T < 1.5 \times 10^{6} K$).

\section{Conclusions}\label{Conclusions}

We have developed a statistical method to search for the populations of weak doublet absorbers in the intergalactic medium that relaxes the requirement that absorbers be  securely identified. We dub this method `agnostic stacking'. In order to characterise the absorber population as a whole, our method is combined with an end-to-end mock analysis testing trial absorber populations in light of our measurements and previously identified systems in the literature.
  
Utilising our stacking and modelling analysis for a set of 26 high-quality COS spectra towards quasars at redshifts z$>$0.7, we aim to detect the signal of \neviii\ absorbers. We arrive at the following results for \neviii\ absorption with median(mean) redshift z$_{abs}$=0.63(0.88): 

\begin{enumerate}
\item  We find a non-detection of both the weak \neviii\ absorber population (12.3 $< \log N \leq$ 13.7), and those with column densities sufficient for direct detection methods (log N$> 13.7$) and place tight limits on the slope and normalization of the CDDF for such absorbers.

\item Combining the results of both search regimes, we are able to place the tightest limits thus far on the \neviii\ absorber population over the full range of column densities accessible to our study (12.3 $<  \log N \leq$15.0). In section \ref{limits_on_CDDF}, we provide an analytic fit to these limits, which are in mild tension with recent predictions from simulations \citep{Rahmati2015}. We also find that there is a 0.024\% probability that the three absorbers along the sightline to PG1148+549 arise at this low velocity separation by chance.

\item We further combined our constraints from stacking with the 13 securely identified strong (log N$_{\neviii } \ge 13.7$) \neviii\ components in the literature for our spectroscopic sample, noting that there is only a 1\% probability that all 13 of these are consistent with our stacking measurement. We favour instead an accounting of the 6 underlying systems associated with these components, for which there is a 4.6 \% probability in light of our stacking constraint. This best fitting model associated with this peak probability is a single power-law CDDF with slope and normalisation, $\beta =-1.86  \substack{+0.18 \\ -0.26}$ and $\log f_{13.7}=-13.99 \substack{+0.20 \\ -0.23}$. The optimally constrained combination of these parameters is $(-\beta ^{0.08}) \times (-\log f_{13.7})^{0.92}$=11.90$\pm$0.29. This leads to a absorber line number density of $dn/dz(z=0.88)=1.38 \substack{+0.97 \\ -0.82}$.  

\item The predicted column density distribution function of \citet{Rahmati2015} is ruled out at between  2.2$\sigma$ and 2.7$\sigma$ (depending on the output redshift used) due to the excess intergalactic \neviii\ produced in their simulations.

\item This column density distribution function can be translated into limits on the cosmological mass density of \neviii, $\Omega  _{\textrm{\neviii}} (z=0.7, 12.3< \log N<15.0) = 2.2 \substack{+1.6\\-1.2} \times 10^{-8}$. This is a factor of three lower than the cosmic density in simulations (Rahmati et al. 2016), and an order of magnitude below the value derived by \citet{Meiring2013} based on their detection of three absorbers along the sightline to PG1148+549. Taking the typical properties of directly detected \neviii\ lines we estimate the  cosmic baryon density associated with the \neviii-bearing gas of $\Omega _{b} \approx 1.8 \times 10^{-3}$, representing 4$\%${} of the total baryon density.

\end{enumerate}

We thank Alireza Rahmati for generously suppyling us with the  CDDFs for \neviii\ from the EAGLE simulations. This work was supported by NASA grant HST-AR-12643.02-A.  
M.M.P was supported by the A*MIDEX project (ANR- 11-IDEX-0001-02) funded by the Investissements d$'$Avenir French Government program, managed by the French National Research Agency (ANR), and by ANR under contract ANR-14-ACHN-0021.

\appendix
\section{Error estimate in the composite spectra}
\label{errors}

The following describes how we estimate the error estimate in the composites at the \neviiib\ position. For a subset of composites, we have looked at four different ways to characterise the noise. First, since there is an absence of absorption features close to \neviiib, we examine the flux distribution in the stack over an interval of $\pm$20 pixels. The error of the mean is one measure of the noise. Additionally, we have fitted this distribution of flux values with a simple Gaussian, and can take the width of the fitted curve as another measure of the noise. Thirdly, we compute the expected noise at the \neviiib\ pixel by adding in quadrature the known noise from each spectrum going into the composite. Lastly, we have also estimated the error by a boot-strap method, in which we constructed different versions of the composite by adding whole spectra (rather than boot-strapping over each pixel's flux distribution). It turns out that all of these four measures yield virtually the same level of noise. Hence, for the rest of the analysis, we take the boot-strap error as our noise estimate $\sigma_{c}${} in the composite.

\section{Upper limits to the \neviii\ absorbers from mocks alone}
\label{limitsfrommocks}
In Section \ref{limits_on_CDDF} and Figure \ref{combined_constraint}{} we refer to an upper limit on the \neviii\ population that can be derived without even stacking the real data, but simply by assessing the amount of additional absorption that is needed in the mock data sets, and comparing this with an estimate that we can obtain from the data themselves in windows deemed from of \neviii\ absorption. Here we detail how this limit is derived.
During the construction of the mock data, for each choice of the \neviii\ absorber population, an additional contribution to the absorption caused by uncorrelated absorbers has to be added. If we compare this additional component with the amount of absorption present already without \neviii\ in the data, and find it too low, we can conclude that the \neviii\ component for this model is producing too much absorption. 
The metric we have chosen here is the mean flux decrement, caused by absorption in all pixels above the flux level we used in order to fit the pixel flux distribution when creating the mocks (F$\geq 0.65$, as detailed in Section 3.1.). We can estimate the mean flux decrement caused by non-\neviii\ absorption by choosing regions in our spectra that are both free of \neviii\ and sufficiently close in redshift and actual wavelength (such that neither redshift evolution effects nor other possible variations (e.g. changes in S/N for different parts of the spectra, or abundance of Lyman series absorbers) are negligible). Here we have chosen to sample in each spectrum the region that corresponds to the restframe wavelengths 785.0 \AA $\leq \lambda \leq$840.0 \AA\ of the putative absorber with the highest possible \neviiia redshift, which is dictated by the individual emission redshift of each QSO and the choice of velocity separation. Once we have this mean flux decrement estimate, FD$_{r}$, and its error in our hands, we compare it to the mean flux decrement caused by the additional absorbers in the mock, FD$_{m}$. 
When FD$_{m}$ + 1$\sigma \leq$ FD$_{r}$ - 1$\sigma$, we rule out the specific model for \neviii\ which thus required too little additional absorption to agree with the data. This is what is represented by the thick black line in Figure 10.

% ***************************************************

\bibliographystyle{mn2e}
\bibliography{references_neVIII_paper}

\bsp

\label{lastpage}

\end{document}